\documentstyle[sprocl,psfig]{article}
\def\PR{\em Phys. Rev.}
\def\MPLA{\em Mod. Phys. Lett. A}

\def\PLB{{\em Phys. Lett.}  B}
\def\PRL{\em Phys. Rev. Lett.}
\def\PRD{{\em Phys. Rev.} D}

\def\be{\begin{equation}}
\def\ee{\end{equation}}
\def\bea{\begin{eqnarray}}
\def\eea{\end{eqnarray}}
\def\cm2{\,{\rm cm}^2}

\def\gtwid{\mathrel{\raise.3ex\hbox{$>$\kern-.75em\lower1ex\hbox{$\sim$}}}}
\def\ltwid{\mathrel{\raise.3ex\hbox{$<$\kern-.75em\lower1ex\hbox{$\sim$}}}}

\begin{document}
\title{On an Incompleteness in the General-Relativistic Description
of Gravitation}
\author{D. V. Ahluwalia}
\address{Physics Division H-846, 
Los Alamos National Laboratory, Los Alamos, NM 87545, USA, and\\
Escuela de F\'{\i}sica, Universidad Aut\'onoma de Zacatecas,
Apartado Postal C-580\, Zacatecas 98068, ZAC., M\'exico}

\maketitle\abstracts{
The recently introduced mechanism of flavor-oscillation clocks has been
used to emphasize observability of constant gravitational potentials
and thereby to question completeness of the theory of general relativity.
An inequality has been derived to experimentally test 
the thesis presented.
}

\section{Introduction}

The gradients of the gravitational potentials are well known to
play a major role in the understanding of motion of the cosmic
bodies. Especially, in the weak-field limit of Einstein's theory of 
gravitation, they are responsible for the description of, say, the
planetary orbits. In contrast to that, importantly, in the same limit, 
there are quantum mechanical effects which depend upon the gravitational 
potentials themselves. For example, it was recently shown that 
in performing a quantum mechanical linear superposition of different mass
eigenstates of neutrinos belonging to different lepton generations,
one may create a so called ``flavor oscillation clock'' 
that has the remarkable property to redshift
precisely as required by the Einstein's theory of
gravitation. \cite{prd98a}

In the present study we demonstrate that such clocks in principle allow
to measure the essentially constant
gravitational potential of the local clusters of the galaxies.
Taken to its
logical conclusion this observation results in the question on the 
completeness of Einstein's theory of gravitation. 

In this communication we systematically explore this question. We come to
the conclusion that, while the
gravitationally induced accelerations vanish in a terrestrial free fall,
the gravitationally-induced phases of the flavor-oscillation clocks do
not. This communication is organized as follows. In the next section 
we define the context of this paper. In Sec. III the 
incompleteness-establishing inequality is derived. Section IV outlines
a possible experiment to the test the inequality. The final Section
contains some concluding remarks and summarizes the presented thesis.

\section{Gradientless Gravitational Potentials}

As is well known, the solar system is embedded in the essentially constant
gravitational potential of the local cluster of the galaxies, the so called
Great attractor. This gravitational potential, denoted by
$\Phi_{GA}$ in the following, may be estimated over the entire solar
system to be\cite{K}
\be
\mbox{Solar system:}\quad\Phi_{GA} \equiv \frac{1}{c^2}\,\phi_{GA}
= - 3 \times 10^{-5}.
\ee

For the present communication
 the precise value of $\Phi_{GA}$ is not
important, but what is more relevant is that it is {\em constant}
over the entire region of the solar system to an exceedingly large
accuracy of 1 part in $R_{GA-S}/\Delta R_{S}$. Here $\Delta R_{S}$
represents the spatial extent of the Solar system, and $R_{GA-S}$
is the distance of the Solar system from the Great attractor.
Taking $\Delta R_S$ to be of the order of Pluto's semi-major axis
(i.e. approximately $40$ AU), and $R_{GA-S}$ to be about $40$ Mpc\cite{K}, 
we obtain $R_{GA-S}/\Delta R_{S}\sim 10^{11}$. For
comparison, the terrestrial  and solar potentials 
on their respective surfaces are of the order 
$
\Phi_{E} = - 6.95 \times 10^{-10}$, $\Phi_S = - 2.12 \times 10^{-6}$,
and therefore much smaller as compared to
$\Phi_{GA}$. Nonetheless, they carry significantly larger gradients over 
the relevant experimental regions.

Yet, the constant potential of the Great attractor that pervades the 
entire solar system is of no physical consequence within the
general-relativistic context (apart from it being responsible for
the overall local motion of our galaxy). Even the parenthetically
observed motion disappears if we hypothetically and   uniformly
spread the matter of the galactic cluster into a spherical mass to
concentrically surround the Earth. Such a  massive shell in its
interior provides an example  of the gradientless contribution to
the gravitational potential that we have in mind.

A terrestrial freely-falling frame which measures accelerations to
an accuracy of less than 1 part in about $10^{11}$ is completely
insensitive to this constant potential. Similarly, since the
planetary orbits are determined by the gradient of the
gravitational potential they too remain unaffected by this
potential. Nonetheless, in what follows we shall show that quantum
mechanical systems exist that are sensitive to $\Phi_{GA}$. The
simplest example for such a system is constructed in performing
a linear superposition of, say, two different mass eigenstates (see Eqs.
(\ref{Fa}) and (\ref{Fb}) below).

In the next section, $\Phi_{GA}$ shall be considered as a physical
and gradientless gravitational potential as idealized in the
example indicated above. This potential   is to be distinguished
from the usual ``constant of integration'' or the ``potential at
spatial infinity.''

\section{An Inequality on the Incompleteness of General Relativity}

%(see Eqs. (\ref{Newton}) and (\ref{Schrodinger}) below). 
In the following we will exploit the weak-field limit 
of gravity as being introduced on experimental grounds.
The phrase `` weak-field limit'' refers to the experimentally 
established limit in the weak gravitational fields, rather than to the 
limit of a specific theory. Further, though not necessary, for the sake of
the clarity of presentation we shall work in the non-relativistic
domain and neglect any rotation that the gravitational source may
have. This assumption shall be implicit throughout this
communication. The arguments shall be confined to the system
composed of the Earth and the Great attractor, and are readily
extendable to more general situations.

\def\gm{\frac{2G M}{c^2 r}}

For the measurements on Earth
the appropriate general-relativistic space-time metric is
\be
ds^2 =g_{\mu\nu}dx^\mu dx^\nu =\left(1-\gm\right)
dt^2-\left(1+\gm\right) d\vec r^{\,2},
\label{gmunu}
\ee
where $M$ is the mass of the Earth, $r$ refers to the distance
of the experimental region from Earth's center, and $d\vec r^{\,2}=
\left(dx^2+dy^2+dz^2\right)$. The
conceptual basis of the theory of general relativity asserts that
the flat space-time metric $\eta_{\mu\nu}$
\be
ds^2 = \eta_{\mu\nu} dx^\mu dx^\nu = dt^2 - d\vec
r^{\,2}\label{etamunu}
\ee
is measured by a freely falling observer on Earth (or, wherever the
observer is). In this framework,  a stationary observer on the
Earth may define a gravitational potential according to 
\be
\phi_E(\vec r)= \frac {c^2}{2}\left(g_{00}-\eta_{00}\right)
=-\frac{c^2}{2}\left(g_{\jmath\jmath}-\eta_{\jmath\jmath}\right),\quad
\jmath=1,2,3\,\,(\mbox{no sum}).\label{phie}
\ee

One immediately suspects that such a description may not
incorporate the full physical effects of such physical potentials
as $\phi_{GA}$ even though this conclusion is consistent with the
classical wisdom. Indeed,
   the classical equation of motion consistent with the
approximation in Eq.~(\ref{gmunu})
\begin{equation}
m_i\frac{\mbox{d}^2 \vec r\,} {\mbox{d} t^2} = -\,m_g\vec\nabla
\phi_E(\vec r), \label{Newton}
\end{equation}
is invariant under the transformation
\be
\phi(\vec r)_E \rightarrow \varphi_E(\vec r)
=\phi_{GA} + \phi_E(\vec r). \label{transformation}
\ee
For this reason $\phi_{GA}$ has no apparent effect on the planetary
orbits.

In the quantum realm the appropriate equation of motion is the
Schr\"odinger equation with a gravitational interaction energy term
\begin{equation}
\left[ -\left(\frac{\hbar^2}{2 m_i}\right)
\vec\nabla^2 + m_g \phi_{grav}(\vec r)\right]
\psi(t,\vec r)
= i\hbar \frac{\partial\psi(t,\vec r)}{\partial t},\label{Schrodinger}
\end{equation}
as has been confirmed {\em experimentally} in the classic neutron
interferometry experiments of Collela, Overhauser, and Werner\cite{COW,JJS}. 
Equation (\ref{Schrodinger})
is not invariant under the
transformation of the type (\ref{transformation}).

Moreover, this lack of invariance does not disappear in
the relativistic   regime where an appropriate relativistic wave
equation, such as the Dirac equation, must be considered. Therefore, the
gravitational potential that appears in Eq. (\ref{Schrodinger})
cannot be identified with $\phi_E(\vec r)$ of Eq. (\ref{phie}).
To treat the contributions from the Great attractor and the
Earth  on the same footing of physical reality, the
following identification has to be made:
\be
\phi_{grav}(\vec r) \equiv \varphi_E(\vec r)
=\phi_{GA} + \phi_E(\vec r).
\ee

A second observation to be made is to note 
that while by setting $m_i=m_g$ in Eq.~(\ref{Newton})
the resulting equation becomes independent of the test-particle
mass, this is {\em not} so for the quantum mechanical equation of
motion (\ref{Schrodinger}).\cite{JJS}

These two distinctions between the classical- and quantum-evolutions
lead to the conclusion that the theory of
general relativity for the description of gravitation 
cannot be considered complete. 
The gravitational potentials as
defined via $g_{\mu\nu}(\vec r)$ carry an independent physical
significance in the quantum realm, a situation that is reminiscent
on the significance of the gauge potential in electrodynamics
as revealed by the Aharonov-Bohm effect. \cite{AB}

The statement on the incompleteness of general relativity is best 
illustrated
on the example of a ``flavor-oscillation clock.''\cite{grf96}
\begin{eqnarray}
\vert F_a\rangle
&=& \cos(\theta) \vert m_1 \rangle + \sin(\theta) \vert m_2
\rangle,\label{Fa} \\
\vert F_b \rangle
&=& -\sin(\theta) \vert m_1 \rangle + \cos(\theta) \vert m_2
\rangle .\label{Fb}
\end{eqnarray}
In the linear superposition of the mass eigenstates we assume ({\em
only} for simplicity) that both $\vert m_1\rangle$ and $\vert
m_2\rangle$ carry vanishingly small three momentum (i.e. are at
rest).

By studying the time-oscillation between the flavor states $\vert
F_a\rangle$ and $\vert F_b\rangle$ one discovers that this system
can be characterized by the flavor-oscillation frequency\cite{grf97}
\be
\Omega^\infty_{a\rightleftharpoons b}
= \frac{\left(m_2-m_1\right) c^2}{2\hbar}.
\ee

The superscript on $\Omega^\infty_{a\rightleftharpoons b}$ is
to identify  this frequency with a clock at
the spatial infinity from the gravitational sources
under consideration (see below).

Now consider this flavor-oscillation clock to be immersed into
the gravitational potential $\varphi_E(r)$.
Then each of the mass eigenstates picks up a {\em
different} phase because the gravitational interaction is of the
form $m \times
\varphi_E(r)$. As a result, one finds that  the new
flavor-oscillation frequency, denoted by 
$\Omega^\prime_{a\rightleftharpoons b}$,  is given by\cite{grf97}
\be
\Omega^\prime_{a\rightleftharpoons b} = \left( 1+
\frac{\varphi_E(\vec r)}{c^2}\right)
\Omega^\infty_{a\rightleftharpoons b}. \label{red}
\ee

This equation is valid for an observer fixed in the global
coordinate system attached to the Earth.

Equation (\ref{red}) would have been the standard gravitational red
shift expression if the $\varphi_E(\vec r)$ was replaced by
$\phi_E(\vec r)$. Freely falling frames $(\cal F)$ do not carry
fastest moving clocks, they carry clocks that are sensitive to
potentials of the type $\phi_{GA}$. A freely falling frame in
Earth's gravity only annuls the gradients of the gravitational
potential while preserving all its constant pieces such as $\phi_{GA}$. 
In denoting
by $\Omega^{\cal F}_{a\rightleftharpoons b}$ the frequency as
measured in a freely falling frame on Earth,  one is led to
\be
\Omega^{\cal F}_{a\rightleftharpoons b} = \left( 1+
\frac{\phi_{GA}}{c^2}\right)
\Omega^\infty_{a\rightleftharpoons b}. \label{redb}
\ee
$\,$ From a physical point of view, $\phi_{GA}$ represents contributions
from all cosmic-matter sources. However, all these contributions
carry the same sign. In addition, in the context of the cosmos,
$\Omega^\infty_{a\rightleftharpoons b}$ becomes a purely
theoretical entity. Nevertheless, as shown below,
$\Omega^\infty_{a\rightleftharpoons b}$ does have an operational
meaning.

As a consequence, the following incompleteness-establishing inequality
is found,
\be{\Omega^{\cal F}_{a\rightleftharpoons b}} <
{\Omega^{\infty}_{a\rightleftharpoons b}}.
\ee
This is the primary result of our communication.

\section{Outline of an Experiment}

To experimentally test the incompleteness of the
general-relativistic description of gravitation and measure the
essentially constant gravitationally potential in the solar system,
we rewrite Eqs. (\ref{red}) and (\ref{redb}) into (to first  order in
the potentials)
\begin{eqnarray}
\frac{\Omega^\prime_{a\rightleftharpoons b}}
{\Omega^{\cal F}_{a\rightleftharpoons b}} &=& 1+
\frac{\phi_E(\vec r)}{c^2},
\label{experimenta}\\
\frac{\Omega^\prime_{a\rightleftharpoons b}}
{\Omega^{\infty}_{a\rightleftharpoons b}}
&=&\frac{\phi_{GA}}{c^2}+\left(1+
\frac{\phi_E(\vec r)}{c^2}\right).\label{experimentb}
\end{eqnarray}
Equation (\ref{experimenta}) shows how the $\phi_{GA}$-dependence
disappears in ${\Omega^\prime_{a\rightleftharpoons b}}/
{\Omega^{\cal F}_{a\rightleftharpoons b}}$. Equation
(\ref{experimentb}), however, indicates that by systematically
measuring ${\Omega^\prime_{a\rightleftharpoons b}}$ as a function
of $\vec r$, e.g. for an atomic system prepared as a linear
superposition of different energy eigenstates, one can decipher
existence of $\phi_{GA}$. Because all terrestrial clocks are
influenced by the same $\phi_{GA}$-dependent constant factor, it is
essential that the flavor-oscillation clocks under consideration integrate the
accumulated phase over different paths, thus probing different
$\phi_E(\vec r)$, and then return to the {\em same} spatial region
in order that all the data interpretation refers to the same time
standard. Such an integration is easily accommodated in Eq.
(\ref{experimentb}). One would then make a  {\em two parameter fit}
in $\{{\Omega^{\infty}_{a\rightleftharpoons b}},
\phi_{GA}\}$  to a large set of the
{\em closed-loop integrated} data on
$\{{\Omega^\prime_{a\rightleftharpoons b}(\vec r)},
\phi_E(\vec r)\}$. Explicitly
\be
\oint_\Gamma
{\Omega^\prime_{a\rightleftharpoons b}} (\vec r) d\ell(\vec r)
=  {\Omega^{\infty}_{a\rightleftharpoons b}}
\left(1+\frac{\phi_{GA}}{c^2}\right)
\oint_\Gamma d\ell(\vec r)
+\frac{\Omega^{\infty}_{a\rightleftharpoons b}}{c^2}
\oint_\Gamma{\phi_E(\vec r)} d\ell(\vec r)\label{experimentc},
\ee
where $d\ell(\vec r)$ is the differential length element along
the closed path $\Gamma$.
By collecting the data on the ``accumulated phase''
$\oint_\Gamma
{\Omega^\prime_{a\rightleftharpoons b}} (\vec r) d\ell(\vec r)$
and the ``probed gravitational potential''
$
\oint_\Gamma{\phi_E(\vec r)} d\ell(\vec r)
$
for a set of $\Gamma$, and fitting a straight line, one may extract
$\{{\Omega^{\infty}_{a\rightleftharpoons b}},
\phi_{GA}\}$. Rigorously speaking, what
one obtains  is  ${\Omega^{\infty}_{a\rightleftharpoons b}}$
and the constant $\phi_{GA}$ as modified by
other cosmic contributions.
Further, these additional contributions may
include {\em extra general-relativistic} contributions
from the yet-unknown interactions that may couple to
the various parameters associated with the superimposed quantum states.

A simple consideration on the magnitude of various gravitational
potentials involved and the accuracy of clocks based on quantum
superpositions of atomic states leads to the tentative conclusion
that the suggested experiment is feasible within the existing
technology. In this regard note is taken that various ionic and
atomic clocks have reached an accuracy of $1$ part in $10^{15}$
with a remarkable long term stability. In addition, workers in this
field are optimistic that a several orders of magnitude improvement
may be expected in the next few years (see, e.g, Barbara  Levi's
recent coverage of this subject in the February 1998 issue of
Physics Today.\cite{clocks})

\section{Concluding remarks and Summary}

In the present study we emphasized observability of the constant
potential of the Great attractor by means of flavor oscillation
clocks. While
in a classical context, the force $\vec F = - m_g \vec
\nabla \phi (\vec r)$ experienced by an object is independent of
gradientless gravitational potentials such as $\phi_{GA}$,
the frequency of the flavor oscillation clocks
depends directly on $\phi_{GA}$ [in addition to
$\phi_E(\vec r)$].

The above considerations
suggest that in a free fall the space-time interval (at least in the
quantum context) is given by
\def\gmf{\frac{2\phi_{GA}}{c^2 }}
\be
ds^2  =\chi_{\mu\nu} dx^\mu dx^\nu=
\left(1+\gmf\right)
dt^2-\left(1-\gmf\right)d\vec r^{\,2}.
\label{chimunu}
\ee
Simultaneously, Eq. (\ref{gmunu}) should be replaced by
\def\gmn{\frac{2\varphi_E(\vec r)}{c^2}}
\be
ds^2 =\psi_{\mu\nu} dx^\mu dx^\nu=\left(1+\gmn\right)
dt^2-\left(1-\gmn\right) d\vec r^{\,2},
\ee
with Eq. (\ref{etamunu}) remaining valid  at ``spatial infinity.''
Such a modification is perfectly justified because of the linearity of the
weak-field limit (where one is able to formulate
the physics in terms of the additive gravitational potentials).

Within the considered framework and approximations, the space-time
curvatures derived from $g_{\mu\nu}$ and $\psi_{\mu\nu}$ are identical.

The reported incompleteness in the theory of general relativity for
the description of gravitation reveals certain similarities 
to the Aharonov-Bohm effect.\cite{AB} Indeed, in the
Aharonov-Bohm effect an observable phase
arises in a region with vanishing field strength tensor
$F^{\mu\nu}(\vec r)$, 
(i.e. in a region with vanishing $4$-curl of the gauge potential
$A^{\mu}(\vec r)$). 
In the effect reported here,  an observable 
phase arises in a region where the contributions of the
$\phi_{GA}$-type constant potentials to the 
curvature tensor
$R^{\mu\nu\sigma\lambda}(\vec r)$ vanish.
Both of the effects mentioned above,
illustrate the circumstance that in quantum mechanical processes
the gauge field $A^\mu(\vec r)$ and the gravitational potential
$g^{\mu\nu}(\vec r)$
may be favored over the corresponding 
fields strength tensor $F^{\mu\nu}(\vec r)$,
and the curvature tensor
$R^{\mu\nu\sigma\lambda}(\vec r)$, respectively.

However, since the number of the independent degrees of freedom
of $A^\mu(\vec r)$ is quite different from that of $g^{\mu\nu}(\vec r)$,
the  analogy between the Aharonov-Bohm effect 
and the one considered here is not complete.

In summary, the local galactic cluster, the Great attractor, embeds
us in a dimensionless gravitational potential of about $- 3 \times
10^{-5}$. In the solar system this potential is constant to about
$1$ part in $10^{11}$. Consequently, planetary orbits remain
unaffected. However, this is not so for the flavor-oscillation
clocks. In a terrestrial free fall  the gravitationally induced
accelerations vanish, but the gravitationally induced phases
of the flavor-oscillation clocks do not. We argued that
there exists an element of incompleteness in the
general-relativistic description of gravitation. The arrived
incompleteness may be subjected to an experimental test by
verifying the inequality derived here.

The origin of the reported incompleteness  lies in the implicit
general-relativistic assumption on  the equivalence of the
space-time metric as measured by a freely falling observer in the
vicinity of a gravitating source  (which in turn is embedded in a
$\Phi_{GA}$-type constant gravitational potential)
 and the space-time metric as
measured by an observer at the ``spatial infinity.''

\section*{Acknowledgments}

This paper represents an evolution of thoughts originally presented 
at the festivities.  After this work was completed George Matsas
(then at the University if Chicago, and now at ITF, Sao Paulo)
showed to me a 1991 page from his research notes\cite{gm}  
which hinted at the 
conclusion now arrived here. I became aware of his considerations 
in the July of 1998 while visiting Sao Paulo.

This essay is adapted from a recent publication of the present author
to honor Professor Sachs.\cite{mpla}

\end{document}